\documentclass{aa}				 
\usepackage{txfonts}
\usepackage{multirow}
\usepackage{graphicx}
\usepackage{hyperref}
\usepackage{natbib}

\bibpunct{(}{)}{;}{a}{}{,}		
\begin{document} 

   \title{Three-dimensional decomposition of galaxies\\with bulge and long bar}

   \author{P. Comp\`ere\inst{1,2}
   			\and
   		   M. L\'opez-Corredoira\inst{1,2}
   			\and 
   		   F. Garz\'on\inst{1,2}
          }

   \offprints{paulc@iac.es}

   \institute{Instituto de Astrof\'isica de Canarias,
   				E-38205 La Laguna, Tenerife, Spain\\
  				\and 
  				Universidad de La Laguna, Dpto. Astrof\'isica, , 
  				E-38206 La Laguna, Tenerife, Spain
           	  }

   \date{Received July 9, 2014; accepted October 13, 2014}
   
 
  \abstract
   {Some observations indicate that the Milky Way has two inner components, a bulge and a long bar, which present a misalignment of $\Delta\alpha\simeq 20º$ that is against the predictions of some theoretical models that are based on numerical simulations.}
   {We wish to determine whether this misalignment between the bar and the bulge can be observed in barred galaxies other than the Milky Way.}
   {Each galaxy of our sample was decomposed based on its $K_s$-band 2MASS image by fitting and modelling in a three-dimensional (3D) space the following components: a disc, a bar, and a bulge. The $\chi^2$ goodness-of-fit estimation allowed retrieving the best-fit angle values for the bar and the bulge to detect any misalignment.}
   {From the 3D decomposition of six barred galaxies, we have detected at least three galaxies (NGC 2217, NGC 3992, and NGC 4593) that present a significant misalignment between the bar and the bulge of $\Delta\alpha > 20º$.}
   {}


   \maketitle


\section{Introduction}

The detailed nature and configuration of the structure of the inner Milky Way is a controversial subject that has provoked some debate in the past decades. The large body of observational data sets of increased sensitivity and spatial coverage, in particular in the NIR bands, for the first time permits discussing this topic with an unprecedented degree of accuracy. The position of the Sun far from the Galactic centre and very close to the Galactic mid plane, where most of the obscuration is concentrated, still makes interpreting the data difficult or somewhat doubtful. 

Analyses of the Milky Way morphology indicate that the centre of our Galaxy
hosts a double-bar type structure:

1) A boxy thick bulge of about 2.5 kpc in length with a position angle (PA) of 15$^\circ $--30$^\circ $ with respect to the Sun-Galactic centre direction \citep[e.g.][]{dwek1995,nikolaev1997,stanek1997,corredoira1997,corredoira2005,rattenbury2007,
cao2013,wegg2013,bobylev2014}, and

2) a long thin bar, an in-plane bar, with a half length of 4 kpc and a position angle of around $45^\circ $ \citep[e.g.][]{weinberg1992,hammersley2000,corredoira2001,corredoira2007,picaud2003,
benjamin2005,cabrera2008,gonzalez2012,amores2013}.

It is assumed that bars form as a result of an instability in deferentially rotating discs \citep{sellwood1981}, whereas bulges are a primordial galactic component. Nevertheless, the misalignment scenario is still subject to some controversy, since the usual theoretical models based on N-body simulations are not able to predict a morphology with these characteristics of a misaligned double component \citep{romero2011,athanassoula2012}. However, without entering the discussion on how well N-body realizations represent the real galaxies, other kinds of analyses contemplate the possibility of a misalignment at some moment of the evolution of bulge+bar \citep{abramyan1986,garzon2014}. While it is clear that N-body simulations are successful in modelling some of the observed features in the bulge-bar structure, thus providing theoretical dynamical support that this is a single structure, there are also observational data that can be more easily reproduced by considering two separate components.

The composed bulge+bar structure with slight misalignment might be caused by the dynamical interaction between a preexisting bulge, as expected in galaxy formation models, and a young bar \citep{valenzuela2009}. \cite{martinez2011} instead proposed a single boxy bulge structure with a twisted major axis to explain the apparent different angle in the in-plane regions. According to \cite{corredoira2011}, the model proposed by \cite{martinez2011} cannot replace the earlier proposal of a bulge+long bar, although the general proposition of an integrated boxy bulge and planar long bar might be possible if a suitable model could be produced that explains all the relevant observational features.

Therefore, our present understanding of the problem is that we apparently see something in the Milky Way for which it is not clear whether it can be explained in terms of theoretical dynamical models. How can this be solved? We must not adopt a deductive standpoint (theorists telling observers what they should see) but an inductive standpoint (deriving theories from observations). To find support for our present conclusions about the Milky Way, we need to find more evidence of this kind of misaligned bulge+bar in galaxies other than the Milky Way to obtain corroboration that this structure is indeed possible: some clear proof that convinces dynamicists to abandon the position that what it is not reproduced by the models cannot be real. This is the aim of this paper: finding some other galaxies with similar morphology.

This evidence was indeed in front of us for many decades. A simple look at the images of some NGC galaxies shows this kind of morphology \citep[e.g.][]{bettoni1994}: a thick bulge and a long bar, in some cases with apparent misalignment. There are some bulge+bar+disc decompositions: \cite{dejong1996}, \cite{castro2003}, \cite{cabrera2004}, \cite{erwin2004} [only with secondary smaller bars], \cite{weinzirl2008}, \cite{gadotti2008}, \cite{kim2014}; 
but since these works applied a two-dimensional (2D) decomposition, extinction and/or projection effects have been proposed as responsible for the apparent misalignment \citep{erwin2013}. This is the point to which we wish to contribute further: by using a 3D decomposition of the components of a galaxy whose possible misalignment cannot be due to perspective because of its inclination, and also using K-band photometry so that the extinction of high-inclination galaxies can be considered negligible. Some similar exercise was also carried out by \cite{erwin2013}, but they focused on features other than the misalignment, as is evident from their analyses.

The paper is organized as follows: in Sect. \ref{data}, we briefly describe how we selected some galaxies and prepared their 2MASS images for our decomposition code. In Sects. \ref{modelling} and \ref{fitting} we present the 3D models of each galactic component and describe the fitting process. We analyse the results in Sect. \ref{results} and compare them with other publications in Sect. \ref{comparison}. Finally, a summary of the results and our conclusions are given in Sect. \ref{conclusions}.


\section{Sample selection and data preparation}
\label{data}

For the purpose of this work, we created a sample of barred galaxies based on a visual inspection of the \textit{Atlas of Galaxies} \citep{sandage1988}. Galaxies with well-defined bar and bulge components without strong morphological perturbations were selected to facilitate and to make the 3D decomposition process more reliable. The NASA/IPAC Extragalactic Database (NED) was used to determine basic characteristics such as the galaxy classification, the inclination and position angles of the galaxy plane, the distance, and the presence of an AGN. AGN are not included in our code. The characteristics of our galaxy sample are shown in Table \ref{selecgal}. The sample list is not meant to be exhaustive because we did not try to include every galaxy in the catalogue that showed the requested features. Instead, our purpose has been to show only a few pertinent examples.

The FITS images used to feed the decomposition code were downloaded from the Two Micron All Sky Survey \citep[2MASS,][]{skrutskie2006} available at the NASA/IPAC Infrared Science Archive webpage (IRSA). We selected for each galaxy the corresponding $K_s$-band image to reduce the interstellar extinction, both internal and in the Galaxy, to prevent fitting problems especially when dust lanes were present in some galaxies of our sample. Because each 2MASS image covers a large 8.5'x17' area with a pixel size of 1"/pixel, we trimmed them around the inner parts of each galaxy to keep mainly the bar and the bulge, cropping almost all of the disc area. We placed the bulge in the centre of the cropped image so that it was compatible with the different coordinate transformation steps of the decomposition routine. To reduce the calculation time, we tried to obtain a small square image of 101 pixels on a side as long as the length of the galactic bar fits in, with the brightest pixel centred on the position $(50,50)$. If some field stars were present in the cropped image, we located and removed them using the $fixpix$ task in IRAF\footnote{IRAF is distributed by the National Optical Astronomy Observatory, which is operated by the Association of Universities for Research in Astronomy (AURA) under cooperative agreement with the National Science Foundation.} because they may induce errors in the decomposition process.

\begin{table}
	\caption{Basic data for the galaxies in our sample.}
		\label{selecgal}
		\centering
		\begin{tabular}{c|ccccc}
			\hline
			\noalign{\smallskip}
			\multirow{2}{*}{Galaxy} & \multirow{2}{*}{Morphology} & $d$ & PA & $i$ & \multirow{2}{*}{Activity} \\
			 &  & (Mpc) & (º) & (º) &  \\ 
			\noalign{\smallskip}
			\hline\hline
			\noalign{\smallskip}
			NGC 1512   & SB(r)a 		& 12.5 & 45 	& 62.0 & AGN 			\\
			NGC 2217   & (R)SB(rs)0/a 	& 19.5 & 104 	& 27.1 & LINER  		\\
			NGC 3351   & SB(r)b 		& 10.2 & 13 	& 53.8 & HII Sbrst 		\\
			NGC 3992   & SB(rs)bc 		& 22.7 & 68 	& 56.6 & LIN. HII 		\\
			NGC 4593   & (R)SB(rs)b		& 30.8 & 100 	& 42.3 & Sey 1 			\\
			NGC 5850   & SB(r)b 		& 20.1 & 139 	& 25.9 & -  			\\
			\noalign{\smallskip}
			\hline
		\end{tabular}
		\tablefoot{Basic characteristics of the selected galaxies according to NED. From left to right: galaxy designation, morphological type, distance $d$ in Mpc, position angle PA, and inclination $i$ of the galactic plane in degrees, finally, the activity classification.}
\end{table}


\section{Modelling the galactic components}
\label{modelling}

After the 2MASS images were correctly prepared for the decomposition process, they were defined in turn as input files for our algorithm to execute a three-step decomposition by fitting first a disc, then a bar, and finally a bulge. We assumed a luminosity proportional to the stellar density for each component. In this section, we describe each of these galactic components our code uses to fit the original image.

\subsection{Disc}

The synthetic disc component is based on a standard thin exponential disc model, that adapts the variation of stellar density in the Milky Way from \cite{corredoira2004}:
%
%
\[\rho_{disc}(R,Z)=
 A\times\exp\left(-\frac{R-R_\odot}{H_{r_1}}-\frac{H_{r_2}}{R}\right)\times \exp\left(-\frac{|Z|}{H_{z}}\right),\]
where the scale lengths $H_{r_1}$ and $H_{r_2}$ are free parameters that our algorithm adjusts during the fitting. Manual tests revealed that changing the scale height $H_{z}$ has a negligible impact on the resulting disc structure, mainly because our selected galaxies appear highly face-on and therefore do not allow a good vertical disc sampling on the reference 2MASS image, even if the fitting algorithm uses a 20 pc vertical resolution. Moreover, the noise level in the 2MASS image is similar to the disc counts and so makes it even harder to precisely determine the $H_{z}$ value of such a faint component. Thus, we decided to fix the scale height parameter to an arbitrary value, choosing that of the Milky Way thin disc: $H_{z}=285$ pc \citep{corredoira2002}. Overall, a precise determination of the disc parameters is not as relevant as for the bar and the bulge parameters because of the faint disc intensity in these inner galactic regions.

\subsection{Bar}

The bar is the second component of the fitting process, adjusted from a previously generated disc-free image. The following equations are used to model a 3D bar based on a 2MASS $K_{s}$-band reference image. The bar density is described by an exponential profile from the galactic centre to the image border:
%
%
\[\rho_{bar}(r_{bar}) = A\times\exp\left (-\frac{{r_{bar}}^2}{s^2}\right ),\]
where $s$ refers to the slope of the exponential distribution, and the elliptical coordinate $r_{bar}$ defines the isodensity contour shape \citep[see][]{corredoira2007} as either elliptical or boxy when $n=2$ or $n=4$, respectively:
%
%
\[
r_{bar} = \left [x_{bar}^n+\left(\frac{y_{bar}}{a_{2,\,bar}}\right)^n+\left(\frac{z_{bar}}{a_{3,\,bar}}\right)^n\right ]^\frac{1}{n},\]
where $a_{2,\,bar}$ and $a_{3,\,bar}$ are the axial ratios of the second and third axis with respect to the major axis of the boxy or elliptical component. The Cartesian reference frame of the bar $(x_{bar},y_{bar},z_{bar})$ is rotated with respect to the disc reference frame by an angle $\alpha_{bar}$ around the Z axis.

For NGC 3992, we manually refined the density profile in the inner region by using this slightly modified expression:
%
%
\[\rho_{bar_{NGC 3992}}(r_{bar})=
 	\left \{\begin{array}{ll}
        		A\times\exp\left (-\frac{{5}^2}{s^2}\right )\,\,\,\,\,\,,& \mbox{$r_{bar} \le 5\, \rm{kpc}$} \\
       			A\times\exp\left (-\frac{{r_{bar}}^2}{s^2}\right )\,\,,& \mbox{$r_{bar} > 5\, \rm{kpc}$}
			\end{array}
	\right \}.
\]

\subsection{Bulge}

As with the bar, the bulge can be modelled based on an elliptical or boxy shape ($n=2$ or $n=4$, respectively). A triaxial model is used not only because it is less restrictive to fit than an axis-symmetric one, but also because most of the bulges show a triaxial structure, starting from the bulge in the Milky Way \citep{corredoira2005} or the Andromeda galaxy \citep{stark1977,gerhard1986,widrow2003}.
%
%
\[
r_{bulge} = \left [x_{bulge}^n+\left(\frac{y_{bulge}}{a_{2,\,bulge}}\right)^n+\left(\frac{z_{bulge}}{a_{3,\,bulge}}\right)^n\right ]^\frac{1}{n},
\]
where its Cartesian coordinates $(x_{bulge},y_{bulge},z_{bulge})$, whose origin is at the galactic centre, are rotated by the bulge angle $\alpha_{bulge}$. Based on \cite{corredoira2005}, the following bulge density expression was used:
%
%
\[\rho_{bulge}(r_{bulge})=
 	\left \{\begin{array}{ll}
        		A\times \left(\frac{r_{cent}}{1\ {\rm kpc}}\right)^{-h}\,\,,& \mbox{$r_{bulge} \le r_{cent}$} \\
       			A\times \left(\frac{r_{bulge}}{1\ {\rm kpc}}\right)^{-h}\,\,,& \mbox{$r_{bulge} > r_{cent}$}
			\end{array}
	\right \},
\]
with $h$ the power-law index of the bulge density and $r_{cent}$ the limit of the very inner region of the bulge where a distribution of higher density is used to better fit the centre of the original image, particularly when a galaxy hosts an AGN. However, since the code does not implement AGN fitting, this free parameter $r_{cent}$ was finally fixed to an intermediate value of 0.35 kpc because the improvement it adds to the fitting is insignificant compared with the much longer execution time it requires when set free to vary. As a rough estimation, fixing this parameter can speed up the fitting by a factor 50 when compared with the case where $r_{cent}$ is free to vary.

After some manual tests, we also decided to fix the bulge shape to elliptical ($n=2$) because of its good agreement with the 2MASS images and the lower residuals it leads to in comparison with a boxy bulge ($n=4$).


\section{Fitting}
\label{fitting}

The aim of the decomposition is to model and fit 3D synthetic disc, bar, and bulge components from a reference image of a given galaxy. For that, free parameters such as ellipticity, length, height, or angle of each galactic component are combined in an iterative way until they converge to an optimal fit when the $\chi^2$ goodness-of-fit indicator is at its minimum. Because our galaxy decomposition code uses a brute-force method, we took advantage of the high-throughput computing system (HTCondor) available at the Instituto de Astrofísica de Canarias (IAC) headquarters to greatly improve the execution time.

To decompose a galaxy, we applied manually defined masks to hide in the 2MASS $K_{s}$-band image all pixels except those we are interested in for the fitting of a given component. It is essential, for example, to mask all the pixels that contribute to the bar and the bulge to fit a disc. This allows the algorithm to converge in a reliable way and run much faster. If we were not to hide all the pixels containing a significant amount of flux from other components than the one we wish to model, the algorithm might be unable to converge and can lead to unreliable fitting.

Our code starts by loading the cropped 2MASS image from which it subtracts the sky background, which is manually determined in IRAF by analysing the original image. The first galactic component to be modelled is the disc. Using masks to select only useful pixels for the fitting, the process generates all the possible synthetic discs in turn, based on the combinations of input values, and projects each of them by taking into account the position angle PA and the inclination $i$ of the galaxy. It is then possible to compare every projected synthetic disc with the corresponding areas of the 2MASS image, quantifying the goodness-of-fit of each modelled disc by computing its $\chi^2$ value. To do so, the elementary disc fluxes are integrated over the lines-of-sight defined by each unmasked pixel and are compared with the corresponding 2MASS image fluxes. The best-fitted disc is detected by retrieving the minimum $\chi^2$ from the analysis of all the synthesized discs. When a valid minimum $\chi^2$ is detected, the best-fitted disc component is modelled using its corresponding parameter values, and a residual \textit{bar+bulge} 2MASS image is created by subtracting the synthesized disc from the \textit{disc+bar+bulge} 2MASS image in such a way that only the bar and the bulge remain, apart from noise present in the original image and low residuals from the recent disc subtraction.

A similar approach is then applied to fit the other two components, a bar and a bulge, using the corresponding residual images. After manually trying boxy and elliptical bars for each galaxy decomposition, we decided for one or the other shape based on the minimum $\chi^2$ value obtained with each one. Two out of six galaxies in our sample, NGC 3992 and NGC 5850, showed better fitting and lower residuals when a boxy bar was used ($n=4$), while the others were best fitted with elliptical bars ($n=2$, see Table \ref{gal_params}). The iterative process then takes advantage of the high-throughput computing system to obtain a precise bar fitting of the predefined shape in a reasonable time. 

Third and last galactic component our code fits, the bulge is faster to model because of its small area and rounded shape. As for the bar, we realized manual tests to determine the best bulge shape. The elliptical shape ($n=2$) was better suited than the boxy one for our selection of galaxies, and so was used during the high-precision bulge fitting. Finally, we obtained a residual image that contains the residual parts of the disc, bar, and bulge subtractions from the original 2MASS file (see those steps illustrated in Fig. 1).

\subsection{Error estimation}
\label{error}

The method used to estimate the uncertainty of the results is derived from the $\chi^2$ value computed for each iteration of our fitting code and the number of free parameters involved in the process. When the fitting of a galactic component was completed, the lowest $\chi^2$ value was retrieved from all the iterations and a 68\% C.L. filter was applied to keep only the best combinations of free parameter values \citep[see][for the table of $\Delta\chi^2$ as a function of confidence level and degrees of freedom, assuming a normal distribution of the sample data, as usual]{avni1976,numericalrecipes}. After this 68\% C.L. selection, the lowest and highest values were retained for every parameter and were carefully examined to detect whether one or more of these values are equal to the limits of the corresponding input ranges. If that was the case, the code had to be executed with wider input ranges until it covered all the valid combinations within that confidence level. Then, the best-fitting combination values were retrieved with their respective errors (see Table \ref{gal_params}).

It appears that some of the disc-parameter uncertainties are meaningless: a significant variation of the disc input parameters sometimes only mildly influences the resulting $\chi^2$ of the modelled disc, mainly because the disc is much fainter than the bar or bulge and in some cases can be of the same order of the background noise, therefore preventing a realistic error estimation. The uncertainty on $\Delta\alpha$, the angle difference between the bar and the bulge, was generally computed within some degrees except for two galaxies of our sample, NGC 1512 and NGC 5850. For NGC 5850, the circular shape of the bulge clearly explains the undefined bulge angle and therefore the important uncertainty on $\Delta\alpha$, but this cannot apply to NGC 1512, which does not seem to have a circular bulge. A possible explanation is a 3D-to-2D projection degeneration: differently shaped 3D components can lead to a similar profile, therefore a similar $\chi^2$ estimation, when they are projected in the 2D sky plane. The same could apply for some highly undefined $a_{3,\,bar}$ or $a_{3,\,bulge}$ values, combined with the general fact that the noise contained in the 2MASS images increased the error estimations.

It is important to note that each of the three galactic components was fitted in a disc-bar-bulge order, each one in a separate process, using the best-fitted result of the previous component. This means that the errors may have been underestimated, mainly for the bulge because it is the last component to be fitted and therefore relies on the best-fitted disc and bar.

To quantify the impact of this sequential approach on the final $\Delta\alpha$ uncertainties, we realized various fittings on the three galaxies in our sample that showed a difference of the bar and bulge angle, $\Delta\alpha$, significantly different from zero: NGC 2217, NGC 3992, and NGC 4593. For each galaxy, we first subtracted the best-fit disc, then created a series of slightly different synthetic bars by varying each bar parameter ($\alpha_{bar}\,$, $a_{2,\,bar}\,$, $a_{3,\,bar}\,$, $s$) within its error margins (Table \ref{gal_params}). The subtraction of each of these bars leads to a series of bulge-only files used by our algorithm to fit the bulge component. The best-fit bulge minimum and maximum parameter values were retrieved to allow for a new error estimation on the bulge fitting. Using this bar-variation method to obtain more reliable error estimations for the bulge angle by taking into account not only the fitting errors, but also the bar-variation errors, we finally obtained the following errors on the bar/bulge angle difference:

\begin{itemize}
\item NGC 2217: $\,\,\,\Delta\alpha_{NGC 2217}=-72.5º^{+5.6}_{-3.3}$
\item NGC 3992: $\,\,\,\Delta\alpha_{NGC 3992}=-63.5º^{+2.3}_{-3.5}$
\item NGC 4593: $\,\,\,\Delta\alpha_{NGC 4593}=+27.0º^{+8.3}_{-7.2}\,\,$.
\end{itemize}

As expected, these error estimates are larger than the fitting-only errors presented in Table \ref{gal_params}, but not to the level that they can affect the general results, especially for well-defined bar models like that of NGC 3992, where the errors are small and therefore neither increase the bulge-parameter nor the $\Delta\alpha$ uncertainties by a significant amount.

Overall, assuming that the shapes of the galactic components obey the formulae given in Sect. \ref{modelling}, it appears that the bar angle and bulge angle, $\alpha_{bar}$ and $\alpha_{bulge}$, are very reliable in that they seem overall well determined, except for some special cases with a circular bulge or when degeneration effects occur, as previously explained.

\subsection{Monte Carlo approach}
\label{montecarlo}

Some decomposition tests were performed using a Monte Carlo method to randomly cover, at the same time, the input domains of the bar and bulge free parameters of NGC 3992. This resulted in a complex fitting with nine degrees of freedom. To reduce the computation time, the best-fitted disc was previously subtracted from the input image, thus eliminating the need to also adjust the disc free parameters that would lead to a problem with 11 d.o.f. But even after our longest computation, which consisted of 100 million iterations calculated by the HTCondor system, the analysis of the best bar+bulge fitting revealed about 30\% more residuals than the results obtained with our sequential one-component-at-a-time approach we described in Sect. \ref{fitting}. Moreover, only 5 combinations out of 100 million were detected within the 68\% C.L. of the best-fit, and 15 within the 95\% C.L., which suggest that even more iterations would be required to achieve a reliable bar+bulge combined fitting, at least in this crude Monte Carlo approach; this is clearly not feasible in reasonable time. The best-fit bar and bulge angles were detected at 127.3º and 11.3º, respectively, and correspond to a minimum of $\chi^2=3690$ computed based on the $N=6962$ pixels used for this combined bar+bulge fitting. These angles are very similar to those measured with our sequential approach (see Table \ref{gal_params}), which confirms the goodness of the sequential approach. However, the number of Monte Carlo iterations needs to be increased to better cover the numerous degrees-of-freedom possible combinations to obtain a reliable fitting when using this combined bar+bulge method.

As previously mentioned, this would be very time consuming, therefore we chose to carry out the galaxy decompositions using a disc-bar-bulge sequence, fitting each component separately by using manually defined masks to select each component area in turn. The fitting then searches for an a priori sufficiently robust solution from which to extract the desired information, while this approach may result on a possible error underestimation, as discussed in Sect. \ref{error}. However, it might be useful, in a future work, to decompose the galaxies by implementing a Markov chain Monte Carlo (MCMC) procedure to explore the parameter degeneracy we can encounter with some bar+bulge configurations by fitting all the components together in an improved way and within a reasonable computation time.


\section{Results}
\label{results}

Table \ref{gal_params} shows for each component of each galaxy in our sample the respective best parameters calculated, which are those that minimize the residuals for every decomposition step. In all cases, the $\chi^2$ goodness-of-fit value is close to the number of pixels, indicating a good agreement of the decomposition with the reference 2MASS image. 

It appears that the eccentricities of the bars in our sample are quite similar for the $a_{2,\,bar}$ ratio, varying from 0.13 to 0.35 with small error margins except for NGC 1512, while the $a_{3,\,bar}$ ratio spreads from 0.08 to 0.90 with much larger uncertainties on the fitting results, especially for NGC 2217, NGC 3351, and NGC 5850. This can be explained by the inclination of the galaxies, where different valid combinations of $a_{2,\,bar}$ and $a_{3,\,bar}$ values in a 3D space can lead to a similar aspect once projected on the 2D sky plane and therefore result in a quite similar low $\chi^2$ value. The same effect can apply for the bar and bulge angles: whereas the error margins of the bulge angle are quite high for NGC 1512, this makes sense because, as said before, different combinations of free parameter values can lead to a similar 2D-projected result while the 3D model has in fact a quite distinct shape. Of course, the high uncertainty on the bulge angle for NGC 5850 is related to its almost spherical shape. 

Overall, the $\Delta\alpha$ values indicate that the bar and bulge angles can be quite different and do not seem to be connected to each other. This leads to the conclusion that the bulge and the bar are not necessarily aligned.

Figure 1 illustrates the fitting and decomposition process on the original image (left-hand column of each galaxy panel), subtracting a synthetic disc, bar, and bulge, which results in a residual-only image that indicates the general fitting quality for each galaxy. When the decomposition is realized, we can combine the synthetic components to recreate the observed 2MASS image. This is shown in the right-hand column, where we display from top to bottom the $disc+bar+bulge$ synthetic image, the $bar+bulge$, and finally the fitted $bulge$ alone. The last residual noisy image displays the difference between the 2MASS original image and the synthetic $disc+bar+bulge$.\\

\subsection{Analysis and comparison with other studies}
\label{comparison}

We estimated the structural parameters of the inner galactic components based on 3D decomposition, while most similar image decompositions are realized in 2D using fitting algorithms such as GALFIT \citep{peng2002,peng2010} or BUDDA \citep{desouza2004,gadotti2008,gadotti2009,kim2014}. It is therefore difficult to directly compare our results with the 2D fits because they use distinct structural parameters, and these 2D fits also focus on other aspects than the angle between the bar and the bulge we are interested in. Moreover, we emphasize the importance of possible problems and weaknesses when using a 2D decomposition for this kind of study. Because various perspective effects and other projection-dependent biases can occur depending on the galaxy inclination and the position angle of each component (see Sect. \ref{error}), it does not seem reliable enough to use 2D fits to obtain reliable deprojected position angles for each galactic component.

\textbf{NGC 1512:} In the centre of its bar, this galaxy hosts a nuclear ring, first noted by \cite{hawarden1979}. Although the 2MASS $K_s$-band image we used to model the components does not explicitly show the inner ring, this kind of peculiar structure might explain part of the high uncertainty on our bulge angle estimate (see Table \ref{gal_params}) because we did not take into account AGN or inner rings during the fitting. \cite{laurikainen2006} realized 2D decompositions that show, for NGC 1512, a difference in the projected bar/bulge angle of $\Delta\alpha=77º$. However, the authors emphasize that they were unable to obtain a reliable decomposition because the image they used was not deep enough, and they had to fit the bar+disc jointly. This means that this galaxy did not allow obtaining reliable results in either 2D or 3D decompositions to conclude on its bar/bulge angle difference.

\textbf{NGC 2217:} Although \cite{desouza2004} noted that their BUDDA 2D decomposition algorithm did not detect any disc in NGC 2217, we had to use a circular disc-like structure in our decomposition to take into account the faint radial gradient visible in the original 2MASS $K_s$-band image. This gradient might be related to dim extensions of the bulge, but our algorithm successfully fitted and removed that faint disc-like component, apart from the bulge, to minimize the total residuals without significantly affecting the bar or bulge parameters because of the much higher flux these two bright components account for. Overall, we cannot compare our 3D decomposition results for NGC 2217 with those reported by \cite{desouza2004} because of the different method -- 2D decomposition -- they used in their BUDDA algorithm.

\cite{jungwiert1997} were the first to detect a double-bar system in NGC 2217. By fitting ellipses, they measured the projected position angles $\rm{PA}_{Bar_1}=112º$ and $\rm{PA}_{Bar_2}=138º$ for the primary and secondary bar. Our decomposition algorithm does not fit the secondary bar, but we can see in Fig. 1 that our NGC 2217 projected bulge angle differs from that of the secondary bar. This indicates that, at least in this case, the secondary bar does not have a significant influence on the bulge fitting because if it had, we would expect the projected synthetic bulge and the secondary bar to be aligned. Moreover, the final residuals are negligible, indicating a good fitting. Therefore, our 3D decomposition appears to be reliable even without fitting the secondary bar. In conclusion, with an angle difference $\Delta\alpha=-72.5º^{+2.5}_{-2.2}$, NGC 2217 seems to effectively present a misalignment between its bar and bulge.

\textbf{NGC 3351:} This galaxy presents a nuclear ring and a hotspot nucleus \citep{curtis1918,sersic1965,buta1996,comeron2010,buta2010,fabricius2012}. Thanks to the masks our algorithm uses to fit each galactic component, these AGN characteristics probably have not affected the disc and bar fitting, but might have affected the convergence of the bulge fitting and therefore the reliability of the bulge angle. On the one hand, a visual inspection or an isophote-based analysis of the original image reveals that the projected bulge appears almost perpendicular to the bar; but on the other hand, the best fit of our 3D decomposition was obtained for a bulge angle similar to that of the bar, meaning that these two components may in fact be well aligned. A perspective effect due to the 53º inclination of the galactic plane might therefore be responsible for the apparent bar/bulge misalignment, although local $\chi^2$ minima were detected during the fitting when the bulge was perpendicular to the bar.

\textbf{NGC 3992:} It is interesting to note the agreement between our results and the analysis based on N-body simulation realized by \cite{erwin2013}. Although their study did not directly focus on the bar/bulge angle, some of their plots clearly show a difference between the bar and the bulge angle in some galaxies, which is perfectly consistent with some of our decomposition results. However, for NGC 3992, which we have in common, although their measured deprojected bar position angle $\Delta\rm{PA}=51º$ agrees with our $\alpha_{bar}=132.5º\pm1.0º$ measurement -- taking into account the 180º symmetry of this component (see Table \ref{gal_params}) --, we observe a significant disagreement with the bulge angle. This has to do with the 2D analysis they made directly on the NGC 3992 image using the isophotes to identify the boxy region, which led to an underestimation of the bar/bulge angle, while our 3D decomposition shows that the angle difference is larger. These are consequences of light-density combinations of each component and projection effects on the 2D sky plane. If we combine our synthetic 3D bar and bulge densities, we obtain a 2D projected pattern similar to the original image (see Fig. 1), which leads to the same isophote-based boxy orientation as measured by \cite{erwin2013}. Overall, NGC 3992 is a clear example of a possible bar/bulge misalignment.

\textbf{NGC 4593:} As demonstrated by \cite{gadotti2008}, the AGN present in this type I Seyfert galaxy significantly affects its inner light profile and therefore the parameters of the 2D decomposition they realized using S\'{e}rsic profile fitting in their BUDDA code.
However, because of the circular symmetry of this feature and the NIR band we used for the analysis, this should not affect our bulge angle study. Therefore, with a difference of $\Delta\alpha=27.0º^{+5.1}_{-6.0}$ (see Table \ref{gal_params}), NGC 4593 is another example of a bar/bulge misalignment. \cite{gadotti2008} realized a decomposition of this galaxy, but because of the 2D approach they used, it is not possible to directly compare their results with our 3D decomposition.

\textbf{NGC 5850:} The best-fit parameter values $a_{2,\,bulge}$ and $a_{3,\,bulge}$ shown in Table \ref{gal_params} indicate that NGC 5850 has a circular bulge. This explains why it is impossible to determine the bulge angle and therefore to conclude about the bar/bulge misalignment. As for NGC 4593, the 2D decomposition made by \cite{gadotti2008} does not allow a direct comparison with our 3D results because of the different methods used.\\

As said, our small sample of six galaxies was analysed to show that, in addition to the Milky Way, there are examples of other galaxies with a thick bulge and a thin long bar, and that a misalignment between the two components is observed in some of them. There are other galaxies that exhibit a similar morphology, for example: NGC 4442 \citep{bettoni1994}, NGC 936, NGC 1433, and NGC 2523 \citep[][Fig 2.]{sellwood1993}; or edge-on S0 galaxies with boxy bulges whose kinematics provide clear evidence for a barred potential, such as NGC 5965 or NGC 5746 \citep{kuijken1996}. Moreover, the thinness of the long bar, $a_{3,\,bar} \ll 1$, was also observed by other authors: \cite{bureau2006} proposed that galaxies with a boxy or peanut-shaped bulge are composed of a thin concentrated disc (a disc-like bar) contained within a partially thick bulge. This thinness of the long bar was also deduced a long time ago from different observations, with typical axes ratios of $a_{3,\,bar}=0.1$  \citep{kormendy1982,wakamutsu1984}, of the order of our rough measurements in our sample of galaxies or in the Milky Way.


\section{Conclusions}
\label{conclusions}

We used the 2MASS $K_s$-band images of six barred galaxies to fit each of them with three synthetic components: a disc, a bar, and a bulge. For this, a decomposition code generated a 3D model for every combination of input parameter values, projected it onto the 2D sky plane by taking into account the position angle and inclination of the original galaxy and then compared the model with the original 2MASS image by computing the $\chi^{2}$ goodness-of-fit indicator. The main results from this decomposition process can be summarised as follows:

   \begin{enumerate}
		\item A given galactic component can be modelled with different shapes in 3D while resulting in a similar profile when projected onto the 2D plane. This degeneracy effect could lead to highly undefined parameter values and therefore might explain the large uncertainties on some results, mainly for $a_{3,\,bar}$ and $a_{3,\,bulge}$.
		\item The bar angle can be quite different from the bulge angle, which indicates that they are not related. Examples from our 3D decompositions are NGC 2217, NGC 3992, and NGC 4593, with an angle difference of up to $72.5º\pm2.5º$ (see $\Delta\alpha$ differences in Table \ref{gal_params}). This would mean that the N-body simulations of barred galaxies need to take into account the possible existence, in the inner part of a galaxy, of at least two independent components that do not have to be aligned.
   \end{enumerate}
   
\begin{acknowledgements}
We would like to thank the anonymous referee for providing us with constructive comments and suggestions. We also would like to thank the language editor for the revision of this paper. This work was supported by the grant AYA2012-33211 of the Spanish Science Ministry. PC was supported by the Spanish FPI funding program of the MINECO under the project AYA2009-06972.

This publication makes use of data products from the Two Micron All Sky Survey, which is a joint project of the University of Massachusetts and the Infrared Processing and Analysis Center/California Institute of Technology, funded by the National Aeronautics and Space Administration and the National Science Foundation.

This research has made use of the NASA/IPAC Extragalactic Database (NED) which is operated by the Jet Propulsion Laboratory, California Institute of Technology, under contract with the National Aeronautics and Space Administration.
      
\end{acknowledgements}

%
%

\begin{table*}[!ht]
	\caption{Best-fit parameters for each galaxy.}
		\label{gal_params}
		\centering
		\renewcommand{\arraystretch}{1.8}	
		\begin{tabular}{cc|c|c|c|c|c|c}
		
			& & NGC 1512 & NGC 2217 & NGC 3351 & NGC 3992 & NGC 4593 & NGC 5850 \\
			\hline
			\hline
			
			\multirow{2}{*}{Disc} & $H_{r_1}$ (kpc) & $11.7^{+\infty}_{-7.2}$ & $4.5^{+0.7}_{-0.5}$ & $15.9^{+\infty}_{-7.9}$ & $4.7^{+0.8}_{-0.6}$ & $6.3^{+2.1}_{-1.3}$ & $8.0^{+2.0}_{-4.3}$ \\
			& $H_{r_2}$ (kpc) & $2.0^{+3.6}_{-\infty}$ & $0.0^{+0.3}_{-0.0}$ & $0.3\pm0.3$ & $16.0^{+1.9}_{-2.1}$ & $4.0^{+1.6}_{-1.5}$ & $1.9^{+3.1}_{-1.1}$ \\
			\cline{2-8}
			& $\chi^{2}$ & 973.84 & 1409.26 & 1043.30 & 856.34 & 2757.19 & 4370.97 \\
			& N & 2116 & 3186 & 2680 & 3239 & 4439 & 3208 \\
			\hline
			\hline

			\multirow{5}{*}{Bar} & $\alpha_{bar}$ (º) & $175.5\pm2.0$ & $9.5^{+0.5}_{-1.0}$ & $101.5^{+2.5}_{-7.0}$ & $132.5\pm1.0$ & $127.5^{+1.0}_{-0.5}$ & $154.0^{+0.5}_{-1.0}$ \\
			& $a_{2,\,bar}$ & $0.19^{+0.21}_{-0.17}$ & $0.35^{+0.04}_{-0.11}$ & $0.31^{+0.04}_{-0.09}$ & $0.27\pm0.02$ & $0.24^{+0.02}_{-0.03}$ & $0.13^{+0.05}_{-0.01}$\\
			& $a_{3,\,bar}$ & $0.16^{+0.07}_{-0.14}$ & $0.23^{+0.37}_{-0.23}$ & $0.90^{+0.10}_{-0.89}$ & $0.27^{+0.03}_{-0.04}$ & $0.08^{+0.33}_{-0.08}$ & $0.27^{+0.73}_{-0.14}$\\
			& $s$ (kpc) & $7.4^{+3.1}_{-1.4}$ & $3.60\pm0.15$ & $2.35^{+1.35}_{-0.20}$ & $6.40^{+0.20}_{-0.15}$ & $10.2^{+0.6}_{-0.7}$ & $14.7^{+0.3}_{-4.5}$\\
			& $n$ & 2 & 2 & 2 & 4 & 2 & 4 \\
			\cline{2-8}
			& $\chi^{2}$ & 1074.80 & 677.10 & 776.84 & 1202.14 & 1566.04 & 2004.90 \\
			& N & 2132 & 1439 & 1304 & 2896 & 2142 & 1981 \\
			\hline
			\hline
			
			\multirow{5}{*}{Bulge} & $\alpha_{bulge}$ (º) & $40^{+36}_{-21}$ & $82.0^{+2.0}_{-2.5}$ & $99.5^{+8.0}_{-6.5}$ & $16.0\pm1.5$ & $100.5^{+6.0}_{-5.0}$ & $135^{+45}_{-135}$ \\
			& $a_{2,\,bulge}$ & $0.73^{+0.05}_{-0.34}$ & $0.81^{+0.02}_{-0.01}$ & $0.92^{+0.02}_{-0.01}$ & $0.43\pm0.03$ & $0.88^{+0.02}_{-0.01}$ & $0.97\pm0.02$ \\
			& $a_{3,\,bulge}$ & $0.67^{+0.15}_{-0.34}$ & $0.12^{+0.39}_{-0.08}$ & $0.48^{+0.03}_{-0.02}$ & $0.75\pm0.02$ & $0.44^{+0.07}_{-0.05}$ & $1.00^{+0.00}_{-0.04}$ \\
			& $h$ (kpc) & $3.18\pm0.06$ & $4.06\pm0.04$ & $3.53^{+0.04}_{-0.03}$ & $2.62\pm0.01$ & $3.15^{+0.02}_{-0.01}$ & $2.89^{+0.02}_{-0.03}$ \\
			& $n$ & 2 & 2 & 2 & 2 & 2 & 2 \\
			\cline{2-8}
			& $\chi^{2}$ & 1224.60 & 3771.29 & 1635.02 & 1848.89 & 5018.98 & 3796.17 \\
			& N & 2504 & 2672 & 1251 & 2815 & 2817 & 2815 \\
			\hline
			\hline
			
			\multicolumn{2}{c|}{$\Delta\alpha=\alpha_{bar}-\alpha_{bulge}$} & $-44^{+21}_{-36}$ & $-72.5^{+2.5}_{-2.2}$ & $2.0^{+7.0}_{-10.6}$ & $-63.5\pm1.8$ & $27.0^{+5.1}_{-6.0}$ & $19^{+135}_{-45}$ \\
			\noalign{\smallskip}
		\end{tabular}
		
		\tablefoot{Results of our 3D decomposition for each galactic component, where $H_{r_1}$ and $H_{r_2}$ are the disc characteristic scale lengths, $\chi^{2}$ and N the minimum chi-square estimate and the number of pixels used for each fitting, $a_{2}$ and $a_{3}$ the axial ratios of the second and third axis with respect to the major axis of the bar/bulge, $n$ refers to the boxy ($n=4$) or elliptical ($n=2$) shape used for the fitting, $\Delta\alpha$ is the difference between the bar angle $\alpha_{bar}$ and the bulge angle $\alpha_{bulge}$. Errors stand only for the fitting of the components separately through the minimization of its $\chi ^2$. We clearly see some cases where the bar and the bulge are not aligned.}
\end{table*}

%
%
\begin{figure*}
		\label{gal_illustrations}
		\centering
		\begin{tabular}{ccc}
			\includegraphics[scale=0.13]{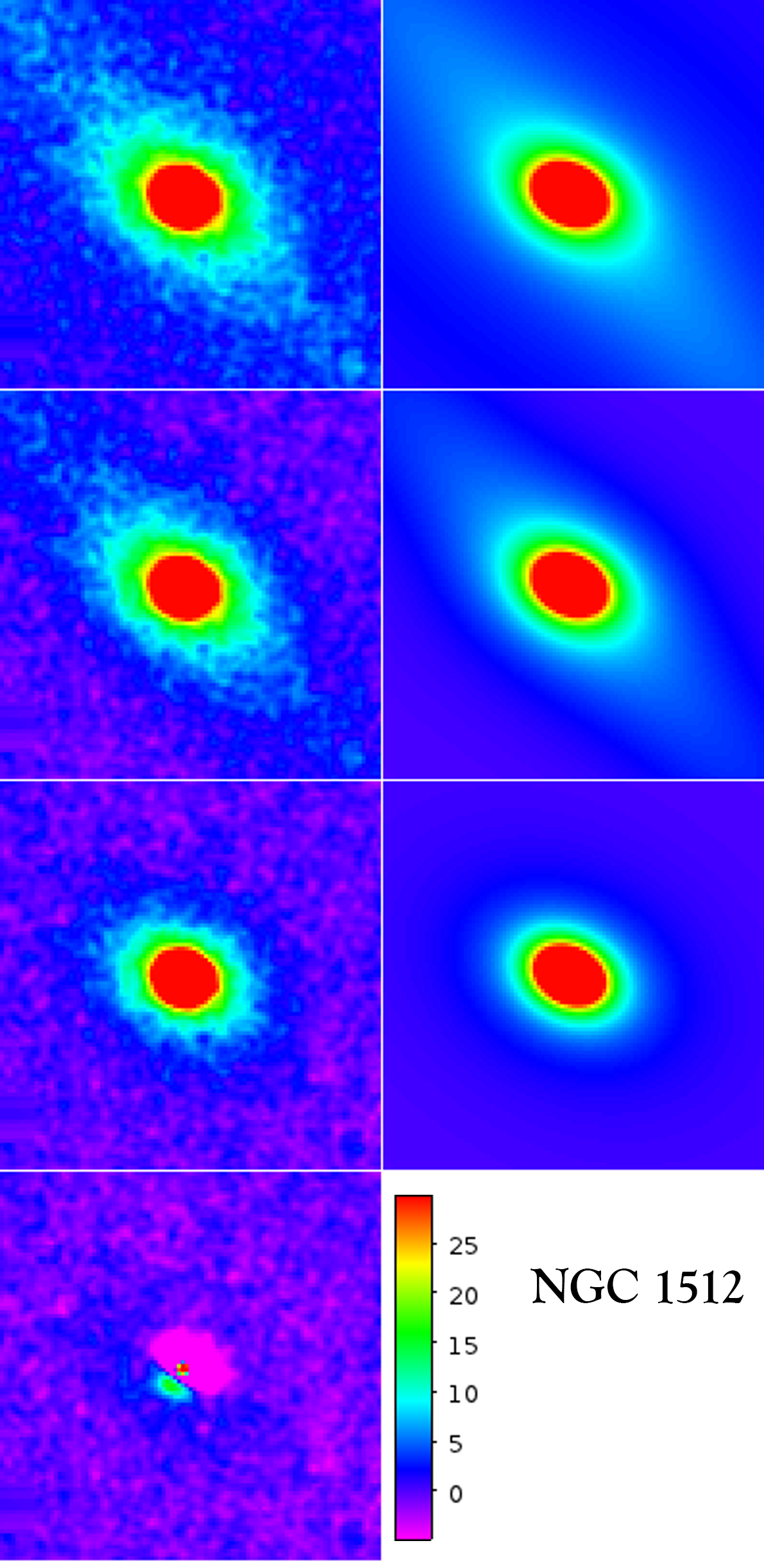} & \includegraphics[scale=0.13]{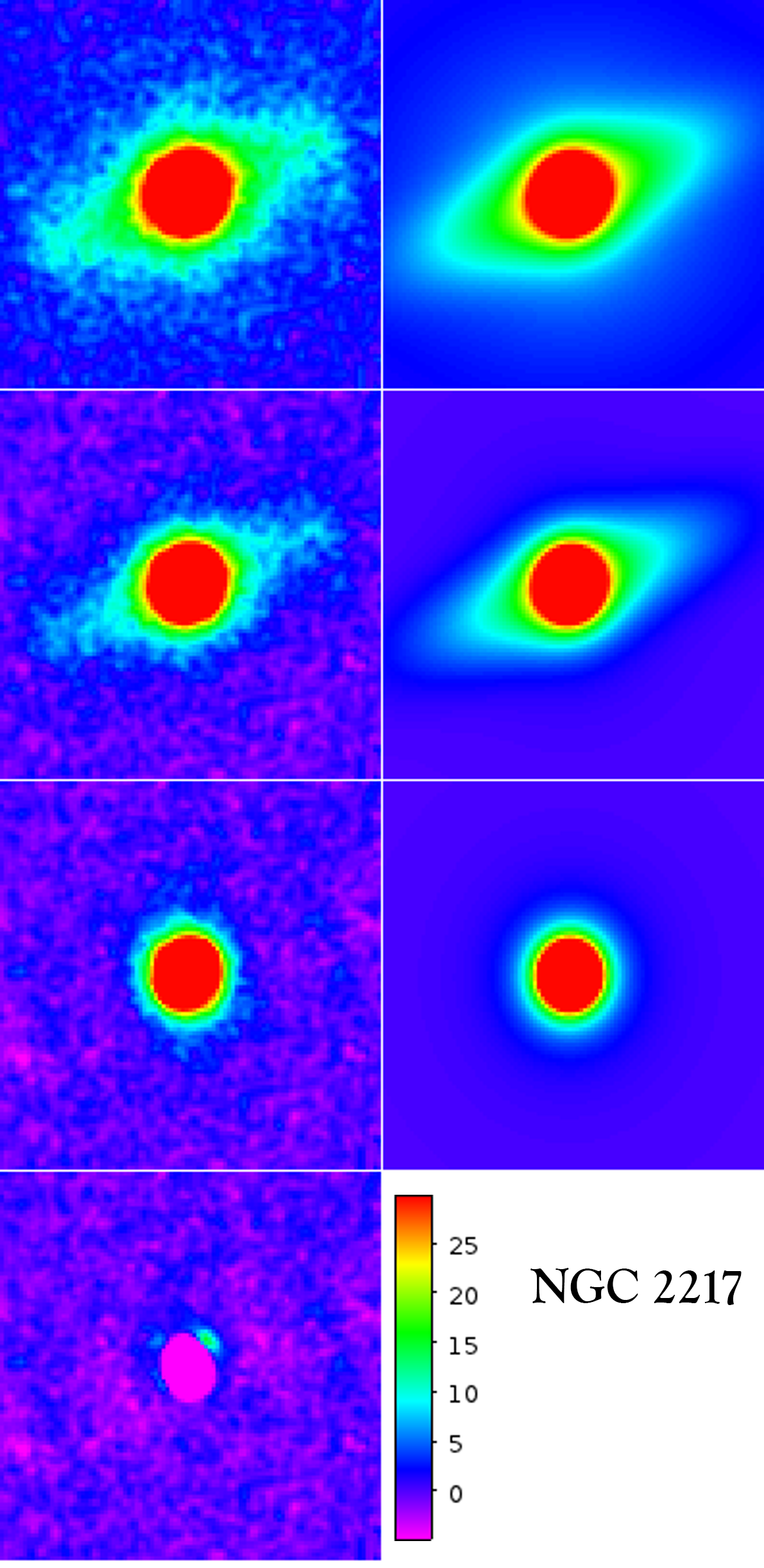} & \includegraphics[scale=0.13]{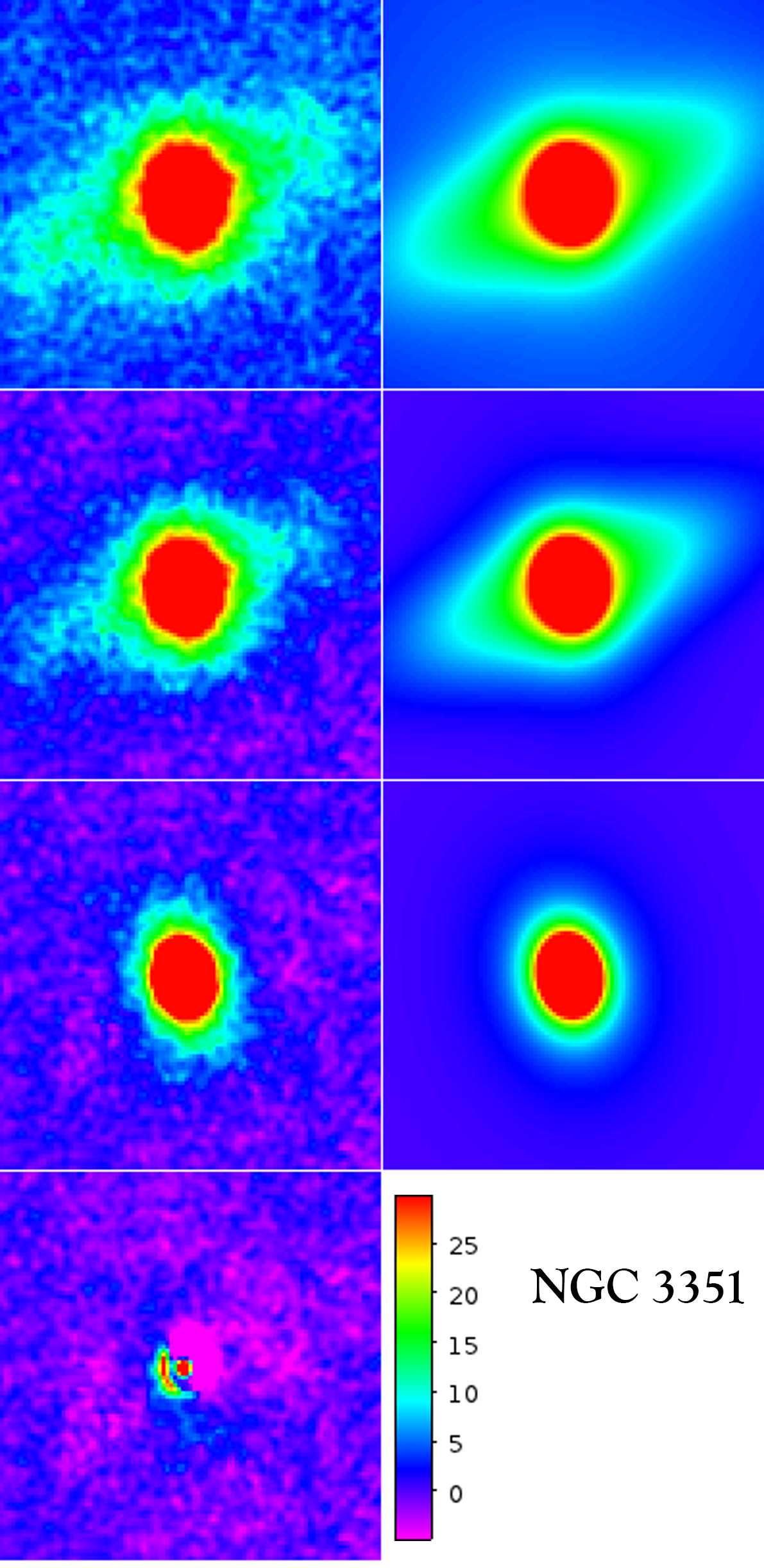} \\
			\noalign{\smallskip}
			\noalign{\smallskip}
			\includegraphics[scale=0.13]{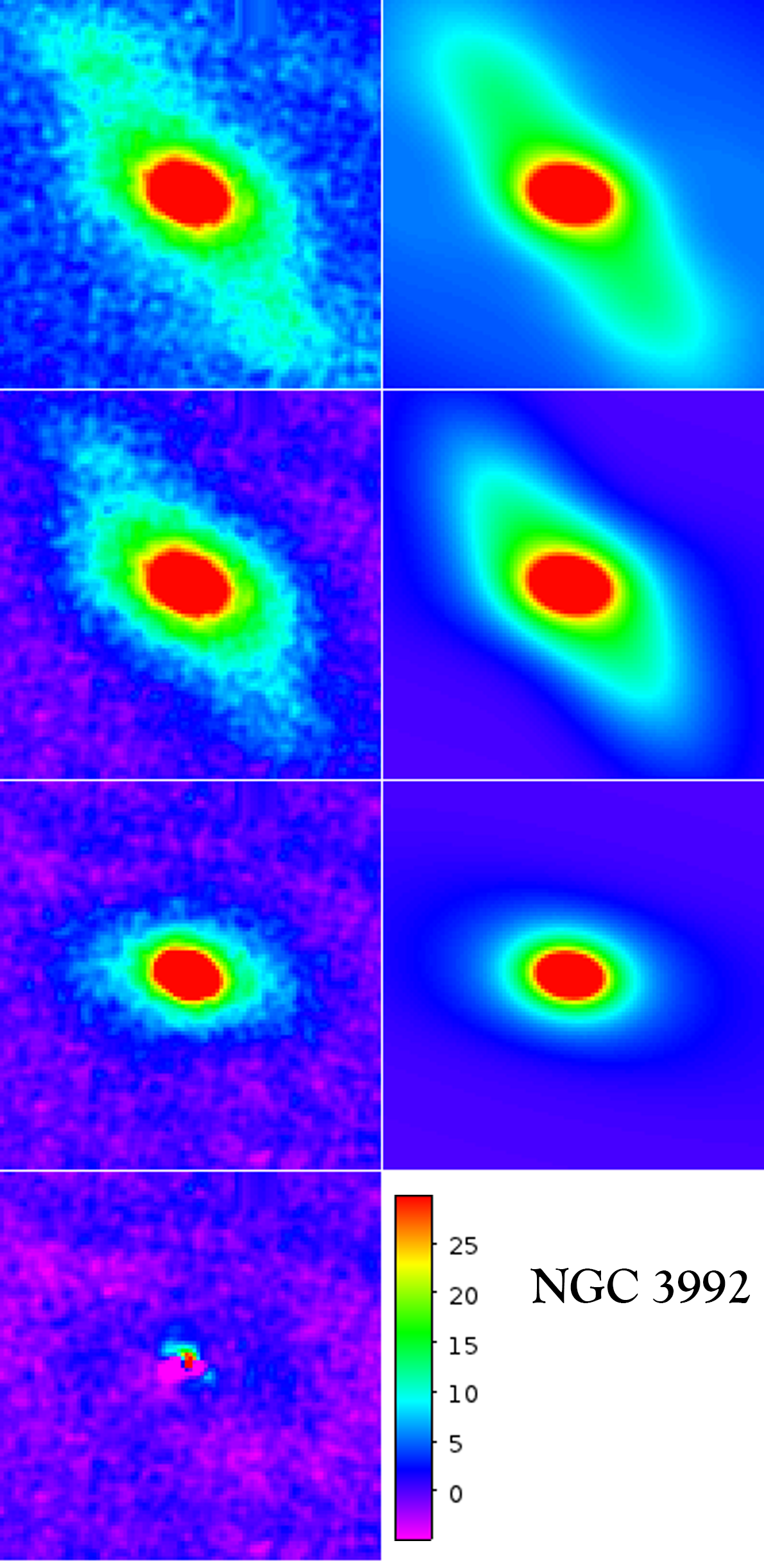} & \includegraphics[scale=0.13]{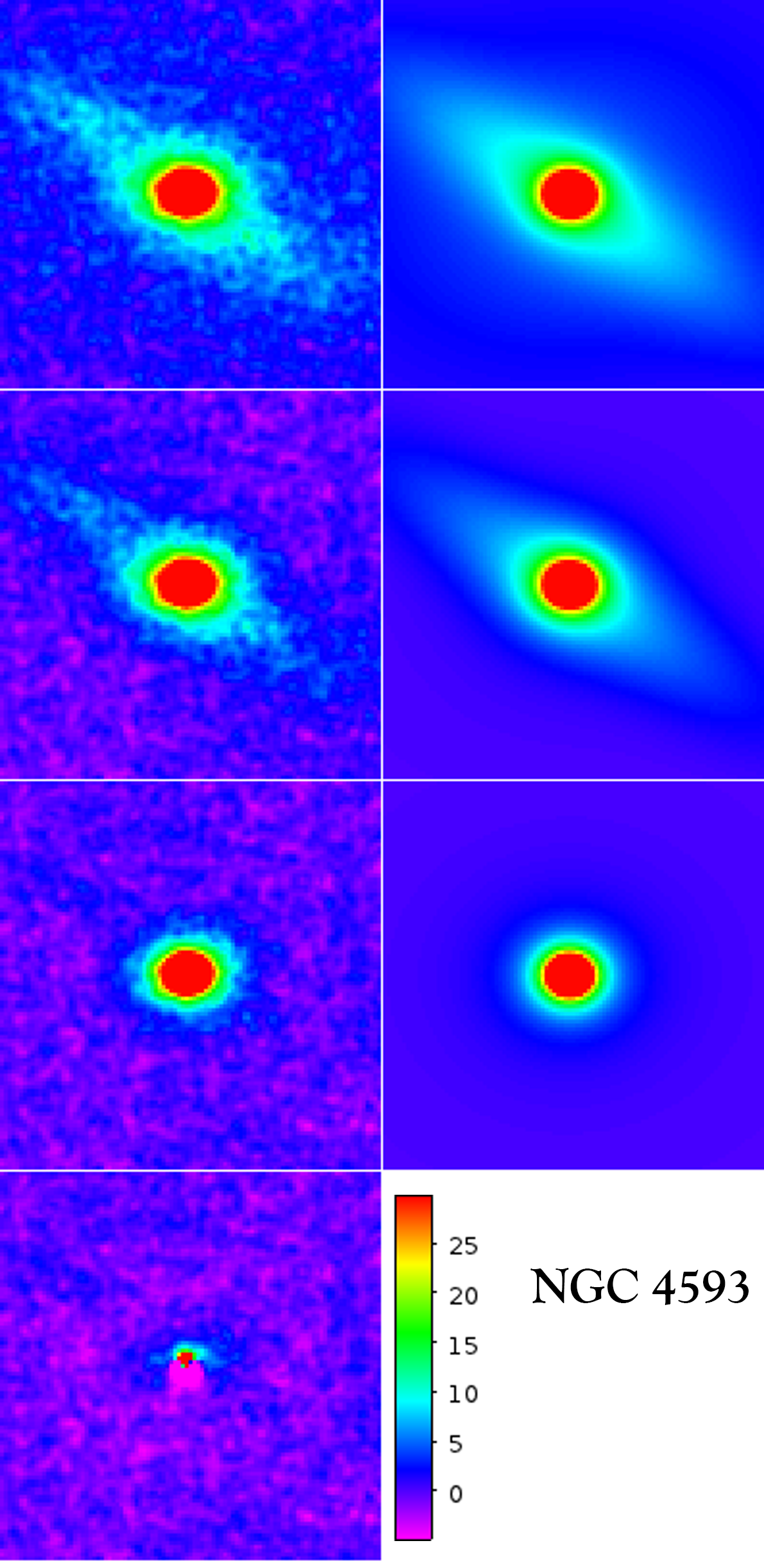} & \includegraphics[scale=0.13]{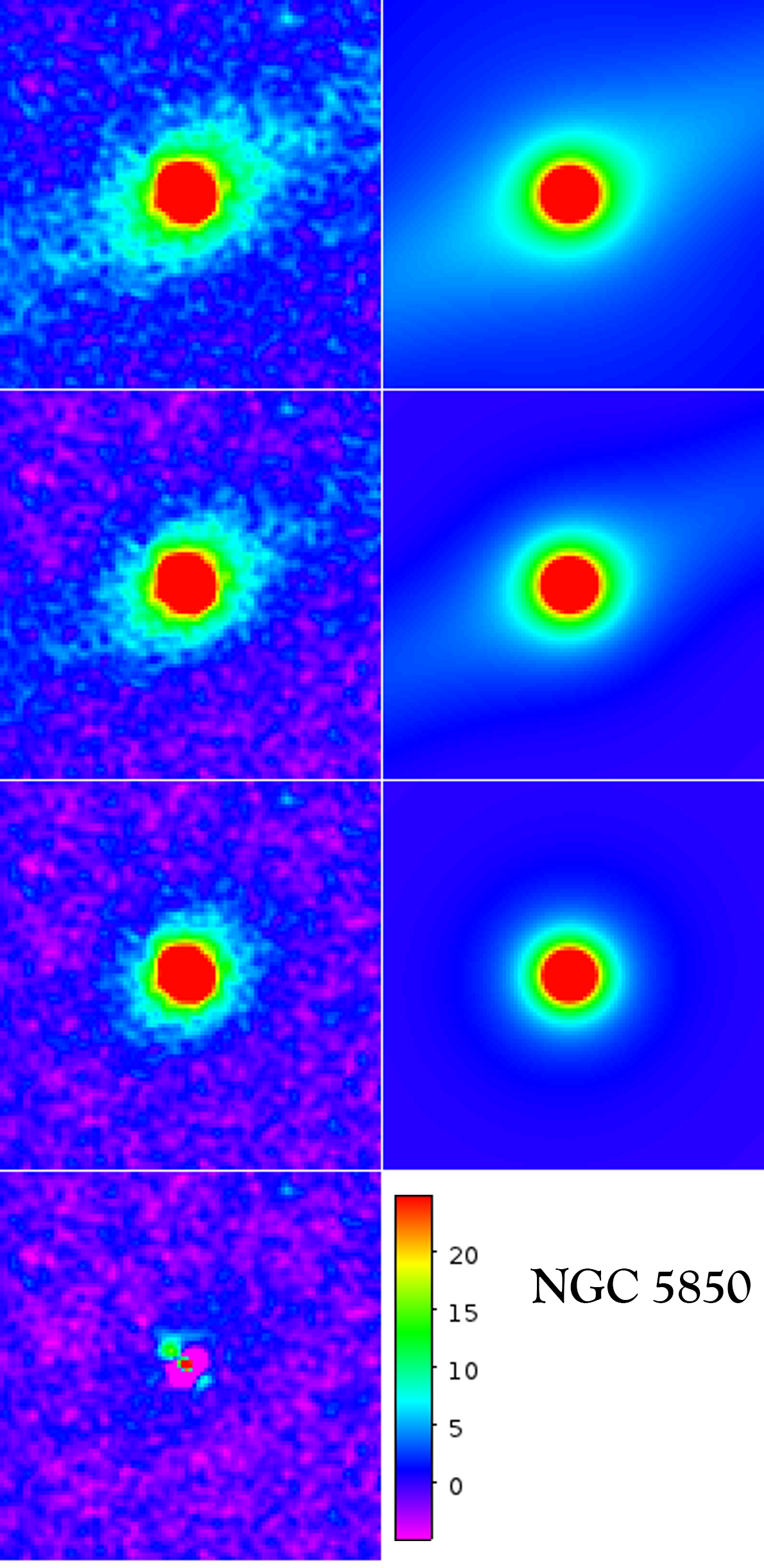} \\
		\end{tabular}
		\caption{Illustration of the decomposition steps and residual images. For each galaxy in our sample each respective panel shows from top to bottom on the left-hand column, the original picture with sky background removed (1), then disc-subtracted (2). After fitting and subtracting a synthetic boxy or elliptical bar (3), the process is completed by modelling an elliptical bulge, trying to minimize the residuals (4). 
		In the right column, the corresponding synthetic data show at the top the fitted $disc+bar+bulge$ image followed by the $bar+bulge$ image, and finally, the synthetic bulge alone. The colour scale units are expressed in ADUs.
		}
\end{figure*}


\nocite{*}
\bibliography{3D_gal_decomp}

\end{document}